\documentclass{svjour2}                    
\smartqed  
\usepackage{graphicx}
\usepackage{amssymb}
\usepackage{color}
\usepackage{mathptmx}      
\usepackage{latexsym}
\begin{document}

\title{Optimal linear Glauber model}

\author{Shaon Sahoo   \and
         Soumya Kanti Ganguly
}

\institute{S. Sahoo \at
               Solid State and Structural Chemistry Unit, Indian
Institute of Science, Bangalore 560012, India\\
              \email{shaon@sscu.iisc.ernet.in}           
           \and
           S. K. Ganguly \at
              Department of Physics, Indian Institute of
Science, Bangalore 560012, India\\
             \email{skganguly@physics.iisc.ernet.in} 
}

\date{Received: date / Accepted: date}

\maketitle

\begin{abstract}
Contrary to the actual nonlinear Glauber model (NLGM), the linear 
Glauber model (LGM) is exactly solvable, although the detailed balance 
condition is not generally satisfied. This motivates us to address the 
issue of writing the 
transition rate ($w_j$) in a best possible linear form such 
that the mean squared error in satisfying the detailed balance condition 
is least. The advantage of this work is that, by studying the LGM 
analytically, we will be able to anticipate how the 
kinetic properties of an arbitrary Ising system 
depend on the temperature and the coupling constants.
The analytical expressions for the optimal values of the parameters 
involved in the linear $w_j$ 
are obtained using a simple Moore-Penrose pseudoinverse matrix.
This approach is quite general, in principle applicable to any system and 
can reproduce the exact results for one dimensional Ising system. 
In the continuum limit, we get a linear time-dependent
Ginzburg-Landau (TDGL) equation from the Glauber's microscopic model 
of non-conservative dynamics. 
We analyze the critical and dynamic 
properties of the model, and show that most of the important results 
obtained in different studies can be reproduced by our new mathematical 
approach. We will also show in this paper 
that the effect of magnetic field can easily be studied within our 
approach; in particular, we show that the inverse of relaxation time changes 
quadratically with (weak) magnetic field and that the 
fluctuation-dissipation theorem is valid for our model. 
\end{abstract}

\section{Introduction}
The nonequilibrium statistical mechanics is a very active field where new 
exciting results are appearing regularly.
While a widely accepted formalism exists for 
studying systems in equilibrium, we are yet to develop a general framework 
to study irreversible processes. 
To get better understanding of how systems evolve, it is 
important to study dynamics of simple physical models. The Ising model is 
probably the simplest non-trivial model in physics. Studying dynamics of 
this model may give us some new insights into the general feature of a 
nonequilibrium process. 

Even for this simple Ising model, studying dynamics is not generally an 
easy task. Roy J. Glauber showed in his classic original work how one can 
study (non-conservative) dynamics in a simple (ferromagnetic) Ising chain 
\cite{glauber63}. Since then this work has been extended to study, both 
analytically and numerically, different systems in numerous physical 
situations with varying degrees of success 
\cite{michael06,grynberg13,uchida07,kong04,godreche00,%
rettori07,gleeson11,goncalves00,stanley87,fisher98}.
For an arbitrary Ising system, it is possible to get a non-linear form of 
the transition rate $w_j$ for which the detailed balance condition is 
exactly satisfied. Unfortunately, 
with this non-linear $w_j$, analytical calculations 
become intractable. It has been a real challenge to develop a microscopic 
kinetic model which will be solvable for a generic system without 
compromising on the detailed balance condition. 

Although the numerical studies of the Glauber dynamics for different Ising 
systems are substantial and satisfactory, the analytical studies, especially 
in the two and three 
dimensions, remain scarce and limited. There are mainly two types of 
analytical approaches to study this microscopic kinetic model: the 
mean field type approaches \cite{leung98,vojta97,puri09,krapivsky10} and the 
quantum formalism for the stochastic models \cite{schad11}. 
The problem of mean field type approaches are 
obvious; they undermine fluctuations. It is though possible to incorporate  
effects of fluctuations by higher order theories, but they naturally come 
with more complexities and one effectively needs numerical methods to study
them. On the other hand, very few system can be exactly solved within the 
quantum formalism. Except for some one dimensional cases (when problems can 
be represented by integrable quantum systems), one needs to use 
different numerical techniques or some approximate methods to study 
dynamics within this approach. In this context, purpose of this paper is to 
present a new analytical approach by which the microscopic dynamics can be 
studied in {\it any} Ising system and which does not undermine fluctuations. 
The microscopic kinetic model that we 
present here is not only exactly solvable in $d$-dimension, it can reproduce 
Glauber's exact results for one dimensional ($d$ = 1) system. In this sense, 
our work can be viewed as the generalization of Glauber's work for one 
dimensional Ising system. We will   
see that, most of the important results regarding the Glauber's dynamics in 
$d$-dimensional Ising system obtained in different studies can 
be reproduced by analyzing our model.

To give a brief formal description about the main idea behind our approach,  
we note that, while the actual nonlinear Glauber model (NLGM) is not 
generally exactly solvable, the detailed balance condition at equilibrium 
is exactly satisfied by the nonlinear choice of transition rate ($w_j$). 
On the contrary, the linear Glauber model (LGM), where the choice of 
$w_j$ is linear, is exactly solvable although the detailed balance 
condition is not generally exactly satisfied 
\cite{scheucher88,oliveira03,hase06}.  
For the LGM, sometimes called the voter model with noise, $w_j$ is taken 
in the following form:
\begin{eqnarray}
\label{w_lgm}
w_j(\sigma_j) = \frac{\alpha}{2}
\left(1-\frac{\lambda}{2d}\sum_{k=1}^z\sigma_j\cdot\sigma_k\right),
\end{eqnarray}
where $z$ in the number of nearest neighbors (or the coordination number; 
$z$ = $2d$ for hypercubic lattice) and $\alpha$ sets the overall timescale of 
the nonequilibrium 
process. The variables $\sigma_k$'s are neighboring Ising spins of the $j$th 
site. The parameter $\lambda$ ($0 \le\lambda \le 1$) determines the strength 
of noise. For one dimensional (ferromagnetic) Ising chain ($d$ = 1), it is 
possible to 
find a $\lambda$ for which the detailed balance condition is exactly satisfied. 
For this special case, $\lambda = tanh~ 2\beta|J|$, where $\beta$ is the 
inverse temperature and $|J|$ is the strength of coupling constant. Unfortunately 
for the two and three dimensional systems, no choice of $\lambda$ satisfies the 
detailed balance condition exactly. 

In this context we pose the following question: For the $d$-dimensional system, 
{\it what is the best value of the parameter $\lambda$ of 
the linear model for which the mean squared error in satisfying the detailed 
balance condition is least?} or in other words, what is the best way of 
relating the parameter $\lambda$ of the LGM to the temperature and the coupling 
constant such that the results obtained from the model are as close as 
possible to the results obtained from the nonlinear Glauber model (NLGM)? 
It is desirable that, the proposed approach to address the issue
should give the exact known value of $\lambda$ for one dimensional system.

In this paper we address the stated issue in a more general set-up. For a 
generic Ising system where a spin is coupled to $z$ neighbors with different 
coupling constants, we show how the nonlinear exact form of the transition 
rate $w_j$ can be linearized in an optimal way. Here obviously, instead of 
one parameter ($\lambda$), there will be $z$ number of parameters in 
general. In our approach, the optimization is done by a 
linear regression process; the Moore-Penrose pseudoinverse matrix involved 
in the regression process is obtained solely from the
configuration matrix and takes a simple form of dimension $z \times 2^z$.
The elements of the pseudoinverse matrix do not depend on the Hamiltonian 
parameters or temperature, this makes our approach very appealing. 
For the obvious reason, we call the present microscopic model 
the {\it optimal linear Glauber model} (OLGM). Here it may 
be briefly mentioned that, though the linear Glauber model was studied 
extensively and still attracts substantial interests, to the best of our 
knowledge, no attempt was made to relate
the parameter $\lambda$ to the temperature and coupling constant for a generic 
Ising system. Therefore ours is the first work in this direction where we 
actually show how to do it in an optimal way.

It is easy to extend our approach to study the effect of magnetic field. 
In this paper we demonstrate that our 
transition rate in the presence of magnetic field works more efficiently 
than the commonly used one. 
As an application, we show that the inverse of relaxation time changes 
quadratically with (weak) magnetic field.
We also discuss the fluctuation-dissipation theorem in the 
present context.

As an additional advantage of the present work, in the continuum limit 
we get a linear time-dependent Ginzburg-Landau (TDGL) equation for 
non-conservative order parameter dynamics. This establishes a connection 
between the phenomenological TDGL theory (linear version) and the Glauber’s 
microscopic model for non-conservative dynamics. 
 
Regarding the nature of the steady state that we obtain from the OLGM, 
we will later see that (section 2.5), although the 
local probability currents are non-zero in the steady state, its average 
over all possible configurations of the neighbors is zero; in addition, 
our approach ensures that the individual opposite currents are on average 
as small in strength as possible. These facts allow us to safely say that, 
within a linearization approach, the steady state that we get here is as 
close to the equilibrium state as possible.  

Our paper is organized in the following way. In section 2, we give a 
detailed description of our approach. In the next section (sec 3), 
we apply our method to study Ising systems in different dimensions. 
We conclude our work in section 4.

\section{General theory}
Let us consider a system of $N$ interacting Ising spins 
($\sigma_i = \pm 1$). These spins can be arranged in any spatial dimension 
where each spin is assumed 
to interact with $z$ neighbors ($z$ is called coordination number). In our 
approach, nature and strength of the coupling constants can be different. 
We here assume that these neighbors are not directly interacting with each 
other, i.e., there is no next-nearest neighbor interactions. Let us now 
consider that $p(\{\sigma\}; t)$ be the probability that the spins take 
the values $\sigma_1, \cdots, \sigma_N$ at time $t$. We note that 
there can be $2^N$ possibilities of spin configurations, and sum of the 
probabilities corresponding to all these possibilities is 1. We now assume 
that $w_j(\sigma_j)$ is the transition rate of $j$th spin, i.e., the probability 
per unit time that the $j$th spin will flip from the state $\sigma_j$ to 
$-\sigma_j$ while the neighboring spins are momentarily remain fixed. This 
rate should intuitively depend on the states of neighbors. We will later 
discuss in detail  
the form of $w_j$'s. Following Glauber, we can now write the master equation 
which gives the time derivative of the probability:
\begin{eqnarray}
\label{mstreq} 
\frac{d}{dt}p(\{\sigma\}; t) = -\sum_j w_j(\sigma_j) 
p(\{\sigma\}; t) + \sum_j w_j(-\sigma_j) p(R_j\{\sigma\}; t),
\end{eqnarray}
where $R_j\{\sigma\}$ represents the same spin configuration as $\{\sigma\}$ 
with $j$th spin flipped.

By considering $\sigma_i(t)$ as stochastic function of time, we can define two 
important quantities, namely, a time dependent average spin value $q_i(t)$ and 
a time dependent correlation function $r_{i,j}(t)$. These are given below, 
\begin{eqnarray}
\label{avq}
q_i(t)& =& \langle \sigma_i(t) \rangle = \sum_{C(N)} \sigma_i 
p(\{\sigma\}; t) \\
\label{avr}
r_{i,j}(t)& =& \langle \sigma_i(t)\sigma_j(t) \rangle = \sum_{C(N)} 
\sigma_i \sigma_j p(\{\sigma\}; t).
\end{eqnarray}
Here sum is over all possible ($2^N$ in number) spin configurations, 
$C(N)$. It may be noted that $r_{i,i} = 1$.

We now write time derivative of these quantities as first step to obtain 
them as function of time. It is easy to get them by multiplying $\sigma_k$ 
and $\sigma_j\sigma_k$ respectively to the Eq. (\ref{mstreq}) and then sum 
them over all possible spin configurations. Following Glauber, these time 
derivatives can be written as, 
\begin{eqnarray}
\label{avq_t}
\frac{d}{dt}q_k(t) =-2\sum_{C(N)} \sigma_k w_k(\sigma_k) 
p(\{\sigma\};t)~~~~~~~~~~\\
\label{avr_t}
\frac{d}{dt} r_{j,k} (t)
= - 2\sum_{C(N)} \sigma_j \sigma_k \{w_j(\sigma_j) 
+ w_k(\sigma_k)\} p(\{\sigma\};t).~~~~~~~ 
\end{eqnarray}

To solve these equations, a choice of $w_j(\sigma_j)$ has to be made. As we 
may expect, the tendency of the $j$th spin to be up or down should depend on  
the states of the neighboring spins as well on the nature and strength of 
the coupling 
constants between $j$th spin and the neighboring spins. For example, if the 
$j$th and $k$th spins are coupled by a ferromagnetic interaction, then the 
$j$th spin will try to align itself parallel to the $k$th spin. There can be 
many ways to choose $w_j(\sigma_j)$ to obey this tendency, 
although the option is constrained by the fact that it has to 
satisfy the equation of detailed balance at equilibrium for all
possible configurations of neighbors. Now we will introduce a general 
mathematical approach by which one will be able to find the optimal linear 
form of the transition rate for an arbitrary Ising system. 

When the system reaches equilibrium at temperature $T$, the probability $p(\sigma_j)$ 
that the 
$j$th spin will be in the state $\sigma_j$ as opposed to $-\sigma_j$ (for a given 
configuration of neighbors), is just proportional to the Maxwell-Boltzmann factor 
$e^{-\beta h_j(\sigma_j)}$. Here $\beta = 1/k_BT$ with $k_B$ being the Boltzmann 
constant. In the factor, $h_j(\sigma_j)$ is the interaction energy associated with 
the $j$th spin (when in the state $\sigma_j$) with its neighbors, and this is 
given by 
\begin{eqnarray}
\label{intenrg}
h_j(\sigma_j) = \sum_{k=1}^z J_k \sigma_j\cdot\sigma_k,
\end{eqnarray}
where $J_k$ is the coupling constant between the $j$th and $k$th spins. For the 
ferromagnetic coupling, $J_k<0$ and for antiferromagnetic coupling, $J_k>0$.
In the 
equilibrium, for a momentarily fixed configuration of other spins, the $j$th spin 
should satisfy the equation of detailed balance,
\begin{eqnarray}
\label{edb}
\frac{p(\sigma_j)}{p(-\sigma_j)}=\frac{w_j(-\sigma_j)}{w_j(\sigma_j)}.
\end{eqnarray}
To proceed further, let us now write the probability factor in the 
following way:  
\begin{eqnarray}
\label{mbprob}
e^{- \beta h_j(\sigma_j)} 
&=&{\rm cosh} \left[\beta\sum_{k=1}^z J_k \sigma_k\right] 
- \sigma_j {\rm sinh}\left[\beta\sum_{k=1}^z J_k \sigma_k\right] \nonumber\\
&=& {\rm cosh} \left[\beta\sum_{k=1}^z J_k \sigma_k\right]
\left\{1 - \sigma_j {\rm tanh}\left[\beta\sum_{k=1}^z J_k \sigma_k\right]\right\}
\end{eqnarray}
If we use the above expression of the probability factor in the equation of 
detailed balance, Eq. (\ref{edb}), we immediately get an exact form of $w_j$, 
\begin{eqnarray}
\label{exctrate}
w_j(\sigma_j) =\frac{\alpha}{2}\left(1+ \sigma_j{\rm tanh}
\left[\beta\sum_{k=1}^z J_k \sigma_k\right]\right),
\end{eqnarray}
where $\alpha$/2 is the transition rate for non-interacting case. This 
nonlinear form of $w_j$ not only fulfills the orientational tendencies of $j$th 
spin mentioned earlier, it exactly satisfies the equation of detailed balance 
(Eq. (\ref{edb})) at equilibrium for all possible configurations. Unfortunately, 
this nonlinear form is intractable for the analytical study of the Glauber 
dynamics. A linear 
form of $w_j$ is easy to handle, but, except for a few special cases, it does not 
exactly satisfy the detailed balance condition. It will be shown here how this 
nonlinear $w_j$ can be linearized in an optimal way such that mean squared error 
in satisfying detailed balance condition is least. It will be also clear in the 
process why we have chosen this particular nonlinear form of $w_j$ while one has 
other options. 

Noting the series ${tanh}~x= x -\frac{x^3}{3} + \cdots$, we can attempt to 
linearize $w_j$ by considering,
\begin{eqnarray}
\label{mpexpr}
{\rm tanh}\left[\beta\sum_{k=1}^z J_k \sigma_k\right] \approx 
\sum_{k=1}^z\gamma_k \sigma_k.
\end{eqnarray}
Here the coefficients $\gamma_k$'s are not just $\beta J_k$'s that appear in the 
first order term of the hyperbolic-tan series. These coefficients also have 
contributions 
from the higher order terms of the series (this will be clear by noting that, 
$\sigma_k^n=1$ if $n$ is even and $\sigma_k^n=\sigma_k$ if $n$ is odd). Although 
by analyzing the series it is possible to find out the exact values of 
$\gamma_k$'s, 
it is best to take the optimized values for the $\gamma_k$'s which can be obtained 
by a linear regression process. By taking the optimized values, we ensure that the 
mean squared error in satisfying the detailed balance condition is least. The 
optimization process somewhat compensates the absence of the nonlinear terms in 
our desired linear form of $w_j$ (nonlinear terms are typically product of 
different $\sigma$'s).
 
To do a linear regression, we will consider $\gamma_k$'s in Eq. (\ref{mpexpr}) as 
the parameters of the regression process. We may note that, Eq. (\ref{mpexpr}) 
actually represents $2^z$ linear equations in $z$ parameters. Each of these linear 
equations corresponds to the one of the $2^z$ configurations of the $z$ neighbors.         
Obviously, no set of values for the $\gamma$'s can simultaneously satisfy the 
overdetermined set of $2^z$ linear equations (except for a special case discussed 
later). We will now see how the best possible values for $\gamma$'s, for which mean 
squared error is minimum, can be obtained.

Before discussing the linear regression process, it may be worth mentioning here 
that, the function $tanh~x$ is linear about the origin ($x = 0$). Since the term 
$[\beta\sum_{k=1}^z J_k \sigma_k]$ is zero or close to zero for a good fraction 
of the total number of configurations (at least for isotropic case when $J_k$'s are 
equal), we expect our linearization to work reasonably good in a normal situation.  

\subsection{Theory of linear regression (LR)}
Before we use it, let us first briefly present the linear regression theory 
\cite{campbell08} necessary for our present work. We consider an overdetermined 
system of $m$ linear equations in $n~(<m)$ unknown coefficients (parameters) 
$\mu_1, \mu_2, \cdots, \mu_n$: $\sum_{j=1}^n a_{ij}\mu_j = y_i$ ($i = 1,2,\cdots,m$). 
Here $y_i$'s are regressands or dependent variables while 
$a_{i1},a_{i2},\cdots,a_{in}$ are regressors or independent variables. This can be 
written in the matrix form as,
\begin{eqnarray}
\label{lnreqn}
A\mu=Y,
\end{eqnarray}
where $A$ is an $m\times n$ matrix with $a_{ij}$ being the $ij$th element, $\mu$ is 
a column vector (dimension $n\times 1$) with $\mu_j$ being the $j$th element and 
$Y$ is again a column vector (dimension $m\times 1$) with $y_i$ being its $i$th 
element. For a particular set of values of $\mu_i$'s, we can define an error 
function $S(\mu) = \sum_{i=1}^m |y_i - \sum_{j=1}^n a_{ij}\mu_j|^2 = ||Y-A\mu||^2$; 
this error function can be minimized with respect to $\mu_i$'s to obtain the best 
possible values for the parameters. It can be easily shown that the minimization 
problem reduced to finding solution of the following equation,
\begin{eqnarray}
\label{eqnmin}
A^TA \mu=A^T Y.
\end{eqnarray}
There are many ways to solve this equation; if $A^+$ be the pseudoinverse matrix of 
$A$, called the Moore-Penrose pseudoinverse, then the solution can be written as, 
\begin{eqnarray}
\label{slsn}
\mu = A^+ Y. 
\end{eqnarray}
If the $n$ columns of the $A$ matrix are linearly independent, then 
$A^TA$ is invertible and the pseudoinverse matrix can simply be obtained as,
\begin{eqnarray}
\label{mpmatrix}
A^+ = (A^TA)^{-1}A^T.
\end{eqnarray}
If $(A^TA)^{-1}$ does not exist, then there are ways to get the pseudoinverse matrix 
$A^+$; for example by doing Tikhonov regularization or by doing a singular value 
decomposition (SVD) of matrix $A$ \cite{campbell08}. We may note that if 
$U\Sigma V^T$ is the SVD of $A$, then $A^+ = V \Sigma^+ U^T$, where $\Sigma^+$ is just 
obtained by taking the reciprocal of each non-zero element on the diagonal of matrix 
$\Sigma$. For our present problem, the $n$ column vectors of the $A$ matrix are  
linearly independent; this is because they are generated by $n$ 
independent Ising variables ($\sigma_i$'s).

We can now use the pseudoinverse matrix to get the minimum of the error function 
$S(\mu)$. Using Eq. (\ref{slsn}) in the error function, we get the following expression 
for the minimum:
\begin{eqnarray}
\label{minerr}
S_{\rm min} = ||(I-AA^+)Y||^2,
\end{eqnarray}
where $I$ is the $m\times m$ Identity matrix. 

\subsection{Application of LR Theory to Glauber dynamics}
To apply this linear regression theory to our problem in hand, we first note that, 
Eq. (\ref{mpexpr}) will give a linear equation in $\gamma$'s corresponding to each of 
the $2^z$ spin configurations. For us $n = z$ and $m = 2^z$. In Eq. (\ref{lnreqn}), 
the elements of the column matrix $\mu$ are the parameters $\gamma_i$'s; we will denote 
this matrix as $\Gamma$. Each of the 
rows of the $A$ matrix will represent one of the $2^z$ spin configurations of the $z$ 
neighbors; we will call this matrix as configuration matrix $C$. For $z = 4$, the 
form of the matrix $C$ can be seen in Eq. (\ref{cz4}).  

\begin{eqnarray}
\label{cz4}
C = \left( \begin{array}{rrrr} 
1 & 1 & 1 & 1 \\
-1 & 1 & 1 & 1 \\
1 & -1 & 1 & 1 \\
1 & 1 & -1 & 1 \\
1 & 1 & 1 & -1 \\
-1 & -1 & 1 & 1 \\
-1 & 1 & -1 & 1 \\
-1 & 1 & 1 & -1 \\
1 & -1 & -1 & 1 \\
1 & -1 & 1 & -1 \\
1 & 1 & -1 & -1 \\
-1 & -1 & -1 & 1 \\
-1 & -1 & 1 & -1 \\
-1 & 1 & -1 & -1 \\
1 & -1 & -1 & -1 \\
-1 & -1 & -1 & -1 \end{array} \right) ({\rm for~~} z = 4) 
\end{eqnarray}

The $i$th element of the column matrix $Y$ is just the value of 
${tanh}\left[\beta\sum_{k=1}^z J_k \sigma_k\right]$ for the spin configuration 
as appears in the $i$th row of the matrix $A$; we will denote this column matrix 
as $\Omega$. With the present set of relevant notations, Eq. (\ref{lnreqn}) is 
rewritten as $C \Gamma = \Omega$ for our regression problem. The best possible 
values of $\gamma_i$'s can be obtained by 
\begin{eqnarray}
\label{gammabest}
\Gamma = C^+ \Omega, 
\end{eqnarray}
where $C^+$ is the Moore-Penrose pseudoinverse of the configuration matrix $C$. 

We will now determine the form of the pseudoinverse matrix $C^+$ for any coordination 
number $z$. We first note that $\sum_{i=1}^{2^z}C^2_{i,j} = 2^z$ for any $j$. We 
also note that, along a row of any two columns $i$ and $j$ ($i \ne j$), there can appear 
only four configurational states, namely, 1 1, 1 -1, -1 1, -1 -1. Each of the states
appears exactly $2^{z-2}$ times; this is due to the fact that, when two Ising spins are 
in one of the four states, rest of the spins ($z-2$ in number) can assume any of the 
$2^{z-2}$ possible configurations. This implies that, 
$\sum_{k=1}^{2^z}C_{k,i}C_{k,j} = 0$, i.e., all the $z$ columns of the $C$ matrix are 
orthogonal. From these facts we can easily conclude that $C^T C$ is a diagonal matrix 
($z\times z$ in dimension) with $(C^T C)_{i,j}=2^z \delta_{i,j}$. This result helps us 
to write the pseudoinverse matrix $C^+$ in the following form,
\begin{eqnarray}
\label{cmpmat}
C^+ = (C^T C)^{-1}C^T = 2^{-z}C^T
\end{eqnarray}

By using this pseudoinverse matrix we can get the best possible values of the 
parameters ($\gamma$'s) from the equation $\Gamma = C^+ \Omega$. This set of 
parameters can then be used in Eq. (\ref{mpexpr}) to obtain a best possible 
linearized version of $w_j$ (see Eq. (\ref{exctrate})),
\begin{eqnarray}
\label{trate}
w_j(\sigma_j) = \frac{\alpha}{2}(1+\sum_{k=1}^z\gamma_k \sigma_j \cdot \sigma_k).
\end{eqnarray}

\begin{figure} []
\includegraphics[width=11.0cm]{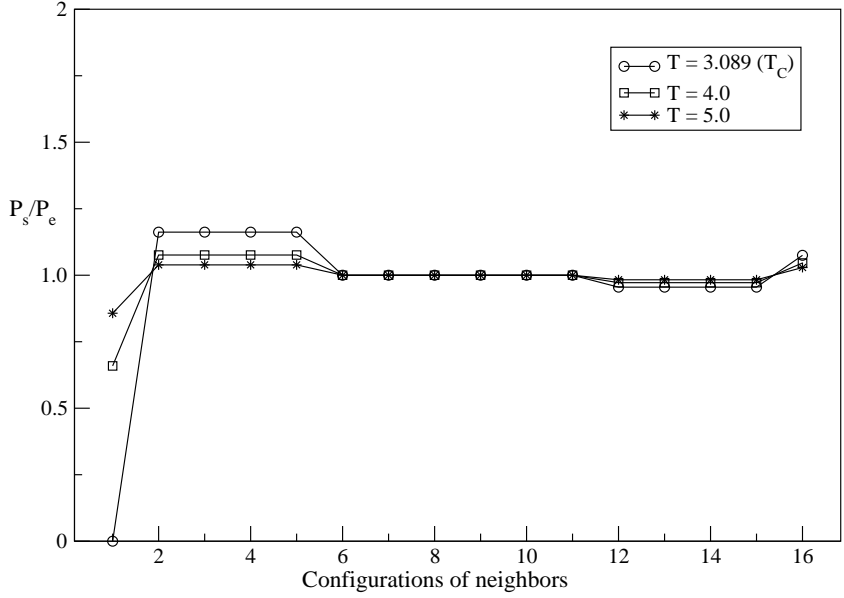}
\caption{Plots show here ratio of the steady state 
probabilities ($P_s$) and 
corresponding equilibrium probabilities ($P_e$) for 16 different 
configurations as given in different rows of $C$ matrix (cf. 
Eq. (\ref{cz4})). Three plots correspond to three different temperatures 
(one corresponds to critical temperature $T_C$ (cf. section 3.2)). The values 
of the temperature are given in the unit of $|J|/k_B$.}
\label{prbplot}
\end{figure}

If we use this linearized form of $w_j$, the net local probability current 
between two spin configurations of a site will not be zero in the steady state 
for the different configurations of neighbors (except for a special case). 
As a result, this steady state does not qualify for an equilibrium state. However, 
our approach makes sure that the average local current is zero and the opposite 
currents are individually as small as possible on the average (cf. section 2.5).  
This allows us to say that the steady state that we get here is as close to the 
equilibrium state as possible within a linearization approach. 
To see how close the steady state probabilities ($P_s$) 
come to the equilibrium probabilities ($P_e$), we plot here ratio of the 
probabilities calculated through our approach and corresponding 
Maxwell-Boltzmann probabilities. The result for the two dimensional system 
can be seen in Fig. \ref{prbplot}, where we have $P_s/P_e = 
[1-\sigma_j |\gamma| \sum_k \sigma_k] / 
[1-\sigma_j~tanh(\beta |J|\sum_k \sigma_k)]$. 
We consider here an isotropic system, therefore we only have one parameter 
$\gamma$ which is given by Eq. (\ref{istrgamma}). In the figure the horizontal 
coordinates represent configurations of neighbors as given in the different 
rows of configuration matrix $C$ (cf. Eq. (\ref{cz4})). We may also note that 
the plots are for a ferromagnetic system ($J = - |J|$), 
and we have taken $\sigma_j= 1$. 
For $\sigma_j= -1$, we will get an exactly similar plots but with a left-right 
inversion (e.g. the value of $P_s/P_e$ corresponding to the 1st configuration 
will now correspond to the 16th configuration in the new figure). 
We see in the figure that 
the ratio is close to 1 for most of the configurations and the value of 
$|P_s/P_e - 1|$ on average decreases with increasing temperature.  

Practical advantage of this linearized form of $w_j$ should be clear now. If we use 
it in Eqs. (\ref{avq_t}) and (\ref{avr_t}), we will get two decoupled sets of 
equations for $q_k$'s and $r_{i,j}$'s respectively. Solving these sets of equations 
is much easier as equations for $q_k$'s will not contain any correlation term 
$r_{i,j}$ and vise versa.

\subsection{Reduction of problem by the use of symmetries} 
The problem of regression can be reduced if 
$Z_2$ symmetry is available, i.e., the external magnetic field is absent. In this case 
we only need to consider half of the configurations which are not transformed to each 
other by $Z_2$ symmetry. The configuration matrix in this case will be of dimension 
$2^{z-1}\times z$; we denote this matrix as $\tilde{C}$. We see that the 
configurations in the first half of the $C$ matrix (1 to $2^{z-1}$ rows) are just the 
spin flipped version of the configurations in the second half. This simply implies that, 
the columns of $\tilde{C}$ matrix are also orthogonal, and consequently 
$\tilde{C}^T \tilde{C}$ is a diagonal matrix with $(\tilde{C}^T \tilde{C})_{i,j} = 
2^{z-1}\delta_{i,j}$. In this case, 
\begin{eqnarray}
\label{cmpmatz2}
\tilde{C}^+ = (\tilde{C}^T \tilde{C})^{-1}\tilde{C}^T = 2^{-z+1} \tilde{C}^T
\end{eqnarray}

Now we will discuss another way of reducing the regression problem. We see that in the 
expression of $w_j(\sigma_j)$ (see Eq. (\ref{trate})), $\gamma_l$ is expected to be same 
as $\gamma_m$ if $J_l = J_m$ in Eq. (\ref{intenrg}), i.e., if $j$th spin is coupled to 
the $l$th and $m$th spin by the same coupling constant. In this case, we can work with 
$z-1$ number of parameters ($\gamma_i$'s) instead of considering all $z$ parameters. We 
also note that, with this consideration, the matrix (which we denote by $\bar{C}$) whose 
pseudoinverse we seek in the regression process, will have $z-1$ columns. It should be 
clear that, one of the columns of the $\bar{C}$ matrix will be just the sum of the $l$th 
and $m$th columns of the actual configuration matrix $C$ matrix, 
and rest of its columns will be same as those of the 
$C$ matrix. Properties of the $\bar{C}$ matrix can be directly derived from the $C$ 
matrix; in fact it can be seen that $\bar{C}^T \bar{C}$ is a $(z-1)\times(z-1)$ diagonal 
matrix with the diagonal elements being $2^z$ except for the row/column which 
corresponds to the sum of two columns of the 
$C$ matrix. The value of this particular element is $2\cdot2^z$ or $2^{z+1}$. 

Now we will consider in little more detail the important special case when all the 
coupling constants are same (the isotropic case). In this case it is enough to consider 
only one parameter ($\gamma$) in the expression of $w_j(\sigma_j)$ 
(see Eq. (\ref{trate})). Clearly, here $\bar{C}$ will be just a column vector and 
consequently $\bar{C}^T \bar{C}$ is just a number whose value is $z\cdot2^z$. For this 
isotropic case, the pseudoinverse matrix (which is a row vector) is, 
$\bar{C}^+ = \frac{2^{-z}}{z}\bar{C}^T$. We may further note that the first element of 
the column vector $\bar{C}$ is $z$, next $^zC_1$ elements are all ($z-2$), and so on 
till we get the last element ($2^z$th) as $-z$. On the other hand, 
the first element of the column vector $\Omega$ is ${tanh}\left[\beta z J\right]$, 
next $^zC_1$ elements are all ${tanh}\left[\beta (z-2) J\right]$, and so on till we 
get the last element ($2^z$th) as ${tanh}\left[\beta (-z) J\right]$. We may note that, 
in this special case, $\Gamma$ will have only one element which is the parameter 
$\gamma$. This parameter $\gamma$ can be calculated using Eq. (\ref{gammabest}); its 
explicit form is given by the following formula:
\begin{eqnarray}
\label{istrgamma}
\gamma = \frac{2^{-z+1}}{z}\sum_{i=1}^{L}Z_i~^zC_{i-1} {\rm tanh}
\left[\beta Z_i J\right], 
\end{eqnarray}
where $Z_i = z-2i+2$ and $L$ is $\frac z2$ or $\frac{z+1}{2}$ depending on whether 
$z$ is even or odd respectively.  
 
It is now worth noting that, for a dimerized linear Ising chain, the coordination number 
$z = 2$. In the absence of magnetic field, we can use the matrix $\tilde{C}$ for the 
regression process. The dimension of this matrix is $2^{z-1}\times z$, i.e., $2\times2$. 
This implies that, we can get exact values of the two parameters ($\gamma_1$ and 
$\gamma_2$) appearing in the expressions of $w_j$'s. This can be also understood from 
the Eq. (\ref{mpexpr}), which in this particular case will give two independent 
equations in two unknown parameters ($\gamma$'s).  When dimerization is zero, these 
two parameters will turn out to be same. This is the special case studied by Glauber 
in his original paper. This shows that our approach to the dynamics is truly general; 
in principle this same approach can be used in studying dynamics of any type of 
Ising system. From this point of view, our approach can be seen as natural extension 
of what Glauber did in his work.  

\subsection{Presence of magnetic field}
In the presence of external magnetic field ($H$), the interaction energy 
associated with the $j$th spin will have now an extra term $-H\sigma_j$, 
and therefore Eq. (\ref{intenrg}) will be modified accordingly,
\begin{eqnarray}
\label{intenrg_h}
h'_j(\sigma_j) = \sum_{k=1}^z J_k \sigma_j\cdot\sigma_k - H\sigma_j.
\end{eqnarray}

It is common to write the modified transition rate ($w'_j$) in terms of  $w_j$ 
in the following way \cite{glauber63}:
\begin{eqnarray}
\label{glbr_trate_h}
w'_j(\sigma_j) = w_j(\sigma_j) (1-\sigma_j{\rm tanh}~\beta H).
\end{eqnarray}

Main problem with this form of $w'_j$ is that, when used in Eq. (\ref{avq_t}), 
the equations for $q_k$'s get coupled with the correlation functions $r_{i,j}$'s. 
On the other side, when this form of $w'_j$ is used in Eq. (\ref{avr_t}), the 
equations for $r_{i,j}$'s not only get coupled with $q_k$'s, they now get 
coupled with more complex three point correlation functions. Here we will 
present another way of handling this issue of applied magnetic field. 
In our approach, the transition rates ($w'_j$'s) will have one more 
independent parameter than 
the number of neighbors ($z$). If we use this form of $w'_j$'s, we will get a 
decoupled set of equations for $q_k$'s ($r_{i,j}$'s do not appear in them) and 
the equations for $r_{i,j}$'s will not contain any three point correlation terms 
(though will get coupled with $q_k$'s). 

To present our approach, we first note that, $w'_j$ will have following exact but 
nonlinear form (cf. Eq. (\ref{exctrate})):
\begin{eqnarray}
\label{exctrate_h}
w'_j(\sigma_j) =\frac{\alpha}{2}\left(1+ \sigma_j{\rm tanh}
\left[\beta \left(\sum_{k=1}^z J_k \sigma_k -H\right)\right]\right),~~
\end{eqnarray}

As before (see Eq. (\ref{mpexpr})), we can attempt to linearize $w'_j$ by 
considering,
\begin{eqnarray}
\label{mpexpr_h}
{\rm tanh}\left[\beta \left(\sum_{k=1}^z J_k \sigma_k-H\right) \right] \approx 
\sum_{k=1}^z\gamma_k \sigma_k +\gamma_H.
\end{eqnarray}

A linear regression will give the best possible values for the parameters 
$\gamma$'s. We here note that,  Eq. (\ref{mpexpr_h}) actually represents
$2^z$ linear equations in $z+1$ parameters. Each of these equations correspond 
to one of the $2^z$  configurations of $z$ neighbors. 

To get the optimal values for the parameters $\gamma$'s, the matrix equation 
to be solved in this case is $C_H \Gamma_H = \Omega_H$ (cf. Eq. (\ref{lnreqn})). 
Here $\Gamma_H$ is a column 
matrix containing $z+1$ parameters ($\gamma_1$, $\cdots$, $\gamma_z$, $\gamma_H$). 
The $2^z\times (z+1)$ matrix $C_H$ is just the configuration matrix $C$ with an 
additional column whose elements are all 1. The column matrix $\Omega_H$ has 
$2^z$ elements, with $i$th element being the value of 
${tanh}\left[\beta(\sum_{k=1}^z J_k \sigma_k-H)\right]$ for the $i$th 
configuration (as appear in the $i$th row of the $C$ matrix). Now 
a linear regression can be done to obtain the best possible values of the 
parameters (for which mean squared error is minimum). 
The best possible values can be formally written using the 
Moore-Penrose pseudoinverse matrix: $\Gamma_H = C^+_H\Omega_H$, here the 
pseudoinverse matrix $C^+_H$ is $(C^T_H C_H)^{-1}C^T_H$. We know that the 
columns of the configuration matrix $C$ are orthogonal, and they are also 
orthogonal to the extra column of the $C_H$ matrix as each column of the $C$ 
matrix has same number of +1 and -1. This implies that the $(z+1)\times(z+1)$ 
matrix $(C^T_H C_H)$ 
is diagonal with the elements being $(C^T_H C_H)_{i,j} = 2^z \delta_{i,j}$. 
Therefore the pseudoinverse matrix in this case is given by, 
$C^+_H = 2^{-z}C^T_H$. As we mentioned, this matrix will help us get best 
possible values of the parameters ($\gamma$'s) from the equation 
$\Gamma_H = C^+_H\Omega_H$. We may here note that all these parameters will 
be the functions of the magnetic field $H$. This optimal values of the set 
of parameters can now be used in Eq. (\ref{mpexpr_h}) to obtain a best 
possible linearized version of $w'_j(\sigma_j)$,
\begin{eqnarray}
\label{trate_h}
w'_j(\sigma_j) = \frac{\alpha}{2}(1+\sum_{k=1}^z\gamma_k 
\sigma_j \cdot \sigma_k+\gamma_H\sigma_j).
\end{eqnarray}

For the important special case when all the coupling constants are same, 
say $J$, it is possible to give explicit formulae for the 
parameters (in the absence of magnetic field, it is given in 
Eq. (\ref{istrgamma})). For this isotropic case, there will be only two 
parameters, $\gamma$ and $\gamma_H$, which are respectively the first 
and second elements of the column vector $\Gamma_H$.
Here the $C_H$ matrix (whose pseudoinverse has 
to be found) has just two columns. All the elements of the second column are 
1. The first element of the first column is $z$, next $^zC_1$ elements of the 
column is ($z-2$), and so on till we get the last element ($2^z$th) of 
the column as $-z$. On the other hand, the first element of the column vector 
$\Omega_H$ is ${tanh}\left[\beta \{zJ -H\}\right]$, next $^zC_1$ elements are 
all ${tanh}\left[\beta \{(z-2)J-H\}\right]$, and so on till we
get the last element ($2^z$th) of the column as 
${tanh}\left[\beta \{-zJ-H\}\right]$.  Since two columns of this reduced 
$C_H$ matrix are orthogonal, $C^T_H C_H$ is a diagonal matrix with first 
diagonal element being $z 2^z$ and second one being $2^z$. This implies that, 
first row of the pseudoinverse matrix $C^+_H$ is just the transpose of the 
first column of  $C_H$ multiplied by $\frac{2^{-z}}{z}$. On the other hand, 
second row of $C^+_H$ is the transpose of second column of $C_H$ multiplied by
$2^{-z}$. Now we can write explicit formulae for the parameters $\gamma$ and 
$\gamma_H$ by using $\Gamma_H = C^+_H\Omega_H$: 

\begin{eqnarray}
\label{istrgamma1_h}
\gamma = \frac{2^{-z}}{z}\sum_{i=1}^{z+1}Z_i~^zC_{i-1} {\rm tanh}
\left[\beta (Z_i J-H)\right]~~{\rm and}\\
\label{istrgamma2_h}
\gamma_H = 2^{-z}\sum_{i=1}^{z+1}~^zC_{i-1} {\rm tanh}
\left[\beta (Z_i J-H)\right],~~~~~~
\end{eqnarray}
where $Z_i = z-2i+2$.

It is easy to verify that, when the magnetic field $H$ becomes vanishingly 
small, the parameter $\gamma_H$ approaches zero. In this limit, the parameter 
$\gamma$ given in Eq. (\ref{istrgamma1_h}) reduces to the parameter obtained 
for the case when the magnetic field was absent (see Eq. (\ref{istrgamma})).

\subsection{Nature of steady state and its closeness to the 
equilibrium state}

It may be noted that, any arbitrary choice of the set of $\gamma$'s would 
make the system evolve to some steady state (which is the solution of the 
master equation at large time), but in general this steady state will not 
be the actual equilibrium state that the given system would relax to. 
That steady state will be the actual equilibrium state of the given system 
only when the equation of detailed balance (Eq. (\ref{edb})) is satisfied 
for all possible spin configurations of the neighbors. Unfortunately, except 
for a special case, no choice of $\gamma$'s 
would obey this condition exactly, as we have less number of parameters than 
the number of configurations (see Eq. (\ref{mpexpr}) and discussion there). 
Our method makes sure that the steady state comes as close to the actual 
equilibrium state as possible within the linearization
approach. Let us see more physically how this is done. First define for the 
$j$th site the net local probability current flowing between two 
configurations, $I_j = w_j(\sigma_j)p(\sigma_j)-w_j(-\sigma_j)p(-\sigma_j)$. 
Here $p(\sigma_j)$ is the the Maxwell-Boltzmann
probability factor defined for the given system while $w_j(\sigma_j)$ is the 
transition rate with an arbitrary set of $\gamma$'s. Clearly the current 
$I_j$ will be positive for some configurations of the neighbors and 
negative for other configurations. 
If we could choose a set of $\gamma$'s for which the equation 
of detailed balance was exactly satisfied for all configurations, then the 
current $I_j$ would have been identically zero for each and every 
configuration. In this context, our method does the following: it 
makes sure that average current $<I_j>$ (average over all possible 
configurations of neighbors) 
is zero in the absence of magnetic field and small, if not 
zero, when magnetic field is present. In addition, our method ensures that 
two opposite tendencies (forward current 
and backward current depending on the sign of $I_j$) are individually as 
low as possible on the average. 

To prove that $<I_j> = 0$ when magnetic field is absent, we first note that, 
this current can be written as, $I_j = \frac{2}{\sqrt{1-X^2}}[Y-X]$; here 
$X = {tanh}\left[\beta\sum_{k=1}^z J_k \sigma_k\right]$ and 
$Y = \sum_{k=1}^z\gamma_k \sigma_k$. We may here note that, in the expression 
of $I_j$, we have missed out a constant prefactor which does not change for 
different configurations. We have also taken $\sigma_j = 1$ to arrive at the 
expression; this 
implies that when $I_j > 0$, there will be a net current flowing from the 
up state to the down state of the $j$th spin (in other words, there will be 
a net tendency for the spin to flip if it's in the up state) and similarly, 
when $I_j < 0$, there will be a net opposite current flowing from the down 
state to the up state of the $j$th spin. Now let us consider two configurations 
(say, $p$ and $p'$) connected by $Z_2$ symmetry. Clearly, if $X_k$ and $Y_k$ 
denotes respectively the values of the $X$ and $Y$ for the $k$th configuration, 
then $X_p = - X_{p'}$ and $Y_p = - Y_{p'}$. This implies that the current 
$I_j$ is exactly opposite for two configurations of neighbors 
related by $Z_2$ symmetry.  
This proves that $<I_j> = 0$ when magnetic field is absent.

In case when magnetic field is present, we again can write the 
current in the form $I_j = \frac{2}{\sqrt{1-X_H^2}}[Y_H-X_H]$, 
where now 
$X_H = {tanh}\left[\beta (\sum_{k=1}^z J_k \sigma_k-H)\right]$ and 
$Y_H = \sum_{k=1}^z\gamma_k \sigma_k + \gamma_H$. Here for two 
configurations, $p$ and $p'$, connected by $Z_2$ symmetry, 
$(X_H)_p \ne -(X_H)_{p'}$ and $(Y_H)_p \ne - (Y_H)_{p'}$. Therefore, 
unlike when magnetic field is absent, the current $I_j$ is now 
not exactly opposite for two configurations connected 
by $Z_2$ symmetry. But we note that, $\sum_{i=1}^{2^z}(Y_H-X_H)_i=0$. 
This can be seen in the following way.
We can write, $\sum_{i=1}^{2^z}(Y_H-X_H)_i = 
\sum_{i=1}^{2^z} (C_H\Gamma_H - \Omega_H)_i$, where 
$(C_H\Gamma_H - \Omega_H)_i$ is the $i$th element of column vector 
$(C_H\Gamma_H - \Omega_H)$ (see section 2.4). We now
have, $\sum_{i=1}^{2^z} (C_H\Gamma_H - \Omega_H)_i = 
\sum_{i=1}^{2^z}\sum_{k=1}^{z+1} (C_H)_{i,k}(\Gamma_H)_k - 
\sum_{i=1}^{2^z}(\Omega_H)_i$. But $\sum_{i=1}^{2^z}(C_H)_{i,k}=0$ for 
all $k$'s except when $k = z+1$, for which
$\sum_{i=1}^{2^z} (C_H)_{i,z+1}(\Gamma_H)_{z+1} = 2^z \gamma_H$. We 
have seen that the value of $\gamma_H$ that we get by regression (or 
by solving $\Gamma_H = C^+_H\Omega_H$) is
$2^{-z}\sum_{i=1}^{2^z}(\Omega_H)_i$. This implies that
$\sum_{i=1}^{2^z} (C_H\Gamma_H - \Omega_H)_i = 0$. 
Noticing that, $|X_H|<1$ for any possible configuration, 
we have the following series, 
$\frac{1}{\sqrt{1-X_H^2}} = 1 +\frac 12 X_H^2+\frac 38 X_H^4 +\cdots$.
In fact when magnetic field and Hamiltonian parameters take 
reasonable values, we expect $|X_H|$ to be well below one for most of 
the configurations. We note that there is no linear term in the above 
series, and variation of existing higher order terms are expected to 
be very small. Therefore as first approximation we take a constant 
value (say, $K$) for $\frac{1}{\sqrt{1-X_H^2}}$. 
This gives, $<I_j> \approx K \sum_{i=1}^{2^z}(Y_H-X_H)_i$, i.e., 
$<I_j> \approx 0$ when magnetic field is present.

We now see that for a linear model, $<J_j>$ is small, if not 
zero, even when the parameters
$\gamma_1$, $\gamma_2$, $\cdots$, $\gamma_z$ (not $\gamma_H$) are chosen
arbitrarily. But arbitrary choice of parameters does not ensure whether the
forward currents and the backward currents are individually as weak as possible.
This is then done by choosing the values of the parameters as obtained by a
linear regression process (see sections 2.1 and 2.2). This facts allow as to 
safely say that, even though our steady state is generally not an equilibrium 
state, it is as close to the equilibrium state as possible within a 
linearization approach.

It is important to study how close we reach to the actual equilibrium state. 
$S_{\rm min}$, as given in Eq. (\ref{minerr}), can be taken as the measure for 
this closeness. Lower the value of $S_{\rm min}$ implies that we are closer to 
the actual equilibrium state. 

In general this minimum of the error function (i.e., $S_{\rm min}$) is not zero 
and its value can be obtained using the pseudoinverse matrix. Taking $C^+$ as 
given in Eq. (\ref{cmpmat}), we get directly from Eq. (\ref{minerr}):  
$S_{\rm min} = ||(I-2^{-z}CC^T)\Omega||^2$. 
Since rows of the $C$ 
matrix are not orthogonal in general, analyzing behavior of $S_{\rm min}$ as 
function of coordination number ($z$) and temperature ($\beta^{-1}$) is not 
always easy. For the special cases we can get simple form of $S_{\rm min}$ 
which allows us to analyze its behavior analytically. 

In the absence of magnetic field, $\tilde{C}\tilde{C}^+$ is an Identity matrix for 
a system with $z = 2$. This implies that, in this special case, $S_{\rm min} = 0$ 
(see Eq. (\ref{minerr})). This result is not unexpected, as we noted earlier, one 
can have exact solution for the $\gamma$'s in this particular situation (or in 
other words, here one does not need to do regression).

Now we will analyze an important special case when all the coupling constants 
are same (say, $J$). 
In the absence of magnetic field, the matrix involved in the regression is a 
column vector ($\bar{C}$) with  $2^{z-1}$ elements. In this isotropic case
the column vector is just the sum of $z$ column vectors of $\tilde{C}$ 
matrix (configuration matrix representing those configurations which are 
not transformed to each other by $Z_2$ symmetry). 
This implies that, the first element of the column vector $\bar{C}$
is $z$, then there are $^zC_1$ number of elements each equals to ($z-2$), and 
so on. We note that, if $z$ is even then the last $\frac 12$$^zC_{\frac z2}$ 
elements of the column vector are all zero. On the other hand, if $z$ is odd 
then the last $^zC_{\frac{z-1}{2}}$ elements of the column vector are all one. 
Following the arguments given in section 2.3, 
it is not difficult to see that, $\bar{C}^T \bar{C} = z \cdot 2^{z-1}$. 
This implies that, the pseudoinverse matrix ($\bar{C}^+$) is a row vector with 
$2^{z-1}$ elements, and is given by $\bar{C}^+ = \frac{2^{-z+1}}{z}\bar{C}^T$.

It is now easy 
to find the elements of the $2^{z-1}\times2^{z-1}$ matrix $(I-\bar{C}\bar{C}^+)$. 
We see that the elements of the first row of the $\bar{C}\bar{C}^+$ is larger 
than the corresponding elements of the other rows. In fact, the elements of the 
first row are $z\bar{C}^+$, the elements of the next $^zC_1$ rows are 
$(z-2)\bar{C}^+$, and so on. Now we note that, 
the first element of $\Omega$ is ${tanh}\left[\beta z J\right]$, next $^zC_1$ 
elements are all ${tanh}\left[\beta (z-2) J\right]$, and so on. It is now not
difficult to see that the absolute value of the first element of the column 
matrix $(I-\bar{C}\bar{C}^+)\Omega$ is the largest.
This allows 
us to get the following upper bound: 
$S_{\rm min} \leq 2^{z-1}|(I-z\bar{C}^+)\Omega|^2$, 
where $I$ is now a row vector whose first element is one and rest of the 
elements are zero. A more appropriate quantity to study here is the minimum 
of the root mean square error ($\bar{S}_{\rm min}$), which is basically the 
minimum of the average error per configuration. This quantity is given by:
$\bar{S}_{\rm min} = \sqrt{\frac{S_{\rm min}}{2^{z-1}}}$. Now using the bound 
for $S_{\rm min}$, we get the following bound for the quantity: 
\begin{eqnarray}
\label{upbnd}
\bar{S}_{\rm min} \leq |(I-z\bar{C}^+)\Omega|.
\end{eqnarray}

We now notice that the absolute value of the first element of the row vector 
($I-z\bar{C}^+$) is larger than the absolute value of any other element. This 
fact is also true for the column vector $\Omega$. This implies that, 
$|(I-z\bar{C}^+)\Omega| \leq 2^{z-1} |(1-\frac{2^{-z+1}}{z} z^2)
{\rm tanh}\left[\beta z J\right]|$. If we use this inequality in 
Eq. (\ref{upbnd}), we get the following upper bound for ${\bar S}_{\rm min}$,
\begin{eqnarray}
\label{rmserr}
\bar{S}_{\rm min} \leq (2^{z-1}- z){\rm tanh}\left[\beta z |J| \right].
\end{eqnarray}

We may here note that, if we had not used $Z_2$ symmetry, we would 
have found twice of what we have got as the upper bound for $S_{\rm min}$; 
the bound for $\bar{S}_{\rm min}$ will though remain same even if we work 
with the full configuration matrix.

Although the upper bound given in Eq. (\ref{rmserr}) is not a tight one, 
but by noting that 
$0 \leq {tanh}\left[\beta z |J|\right] \leq 1$, it makes some sense to 
infer the followings. $\bar{S}_{\rm min}$  is a weak function of temperature 
($\beta^{-1}$) and coupling strength $|J|$; in fact, $\bar{S}_{\rm min}$ 
increases slowly with the parameter $\beta|J|$. Though Eq. (\ref{rmserr}) 
suggests a strong dependence of $\bar{S}_{\rm min}$ on the coordination 
number $z$, we will see in the following sections that the result (critical 
temperature) obtained for three dimensional system is somewhat better than 
that for two dimensional system.

A similar bound can be found in case of non-zero external magnetic field. 
Without going into details of calculation, we can safely say that 
$\bar{S}_{\rm min}$ is a weak function of magnetic field $H$, as it 
is of the parameter $\beta|J|$. This is due to the fact that $H$ appears 
in $\bar{S}_{\rm min}$ only through the argument of hyperbolic-tan function.  

\section{Application to Ising systems in different dimensions}
In this section we will study different Ising spin systems using the method 
we developed in the preceding section. In particular we study the relaxation 
time both in the presence and absence of magnetic field. From the divergence 
of this relaxation time, it is possible to estimate critical temperatures of 
different systems. 
We may here note that, the analysis of our optimal 
linear Glauber model (OLGM) and the linear Glauber model (LGM) are essentially 
same; the advantage of our present work is, we will now 
get to know how the static and dynamic properties of an Ising model depend on 
temperature and coupling constants. We have added a subsection (3.3) to discuss 
different scaling properties of a linear model in the present context. 

\subsection{Relaxation time for a generic system}
For a general lattice, the transition rate $w_j$ in Eq. (\ref{trate}) is 
rewritten  as,
\begin{eqnarray}
\label{trate_lt}
w_{\vec{r}}(\sigma_{\vec{r}}) = \frac{\alpha}{2}(1+\sum_{\vec{R}}
\gamma_{\vec{R}}\sigma_{\vec{r}} \cdot \sigma_{\vec{R}}).
\end{eqnarray}
Here $\vec{r}$ is the position vector of a site while $\vec{R}$ is the 
separation vector identifying neighbors connected to the site. In this 
notation, the equation for $q_k$ as given in Eq. (\ref{avq_t}) is recasted 
as,
\begin{eqnarray}
\label{avq_lt}
\frac{d}{dt}q_{\vec{r}}(t) =-2\sum_{C(N)} \sigma_{\vec{r}} w_{\vec{r}}
(\sigma_{\vec{r}}) p(\{\sigma\};t)~~~~~~~~~
\end{eqnarray}

Now if we use Eq. (\ref{trate_lt}) in Eq. (\ref{avq_lt}), we will get the 
following equation for $q_{\vec{r}}$,
\begin{eqnarray}
\label{avq_lt1}
\frac{d}{dt}q_{\vec{r}}(t) =-\alpha\left[q_{\vec{r}}(t)+
\sum_{\vec{R}}\gamma_{\vec{R}}q_{\vec{R}}(t)\right]. 
\end{eqnarray}

If we denote the total magnetization, $\sum_{\vec{r}} q_{\vec{r}}(t)$, by $M(t)$,
then it is not difficult to see that, $\sum_{\vec{r}} q_{\vec{R}}(t) = M(t)$ for 
a particular $\vec{R}$. To get the equation for $M(t)$, we sum both sides of the 
Eq. (\ref{avq_lt1}) over all sites; this gives,
\begin{eqnarray}
\label{mag_eq}
\frac{d}{dt}M(t) =-\alpha\left[1+
\sum_{\vec{R}}\gamma_{\vec{R}}\right]M(t). 
\end{eqnarray}

Solution of this equation gives us,
\begin{eqnarray}
\label{mag}
M(t) =M(0)e^{- t/\tau},
\end{eqnarray}
where $M(0)$ is the magnetization of the system at $t=0$, and 
\begin{eqnarray}
\label{rlxt}
\tau = \frac{1}{\alpha}\left[1+\sum_{\vec{R}}\gamma_{\vec{R}}(T)\right]^{-1}
\end{eqnarray}
is the relaxation time of the system at temperature $T$. In the above 
expression it is explicitly shown that the $\gamma$'s are all functions 
of $T$.

When magnetic field is present, the transition rate $w_{\vec{r}}$ has to 
be replaced by appropriate $w'_{\vec{r}}$ (see Eq. (\ref{trate_h}) where 
$w'_j$ is given). A simple calculation now leads us to the following 
equation for the magnetization,
\begin{eqnarray}
\label{magh_eq}
\frac{d}{dt}M(t) =-\alpha\left[1+
\sum_{\vec{R}}\gamma_{\vec{R}}\right]M(t) -\alpha N \gamma_H, 
\end{eqnarray}
where, $N$ is the total number of sites in the system. Solution of this 
equation gives us (assuming $H$ is time independent),
\begin{eqnarray}
\label{magh}
M(t) =M(0)e^{- t/\tau_H} - \alpha N\tau_H\gamma_H(1-e^{- t/\tau_H}),
\end{eqnarray}
where $M(0)$ is the magnetization of the system at $t = 0$, and
\begin{eqnarray}
\label{rlxth}
\tau_H = \frac{1}{\alpha}
\left[1+\sum_{\vec{R}}\gamma_{\vec{R}}(T,H)\right]^{-1}
\end{eqnarray}
is the relaxation time of the system at temperature $T$. In the above
expression it is explicitly shown that the $\gamma$'s are all functions
of $T$ and $H$.
We may note that, while in the absence of magnetic field, the system 
relaxes to a non-magnetic/paramagnetic steady state 
($M(t\rightarrow \infty) = 0$) with time scale $\tau$, in the presence of 
uniform magnetic field $H$, the system relaxes to a magnetic steady state 
($M(t\rightarrow \infty) = - \alpha N\tau_H\gamma_H$) with time scale 
$\tau_H$.

To know how the relaxation time changes with magnetic field, we first note 
that, $\tau_H^{-1} = \alpha (1+ z \gamma)$ for isotropic system 
(cf. Eq. (\ref{rlxth})), where the 
parameter $\gamma$ is given by Eq. (\ref{istrgamma1_h}). It is not difficult 
to check that, $z \frac{\partial \gamma}{\partial H}|_{H=0} = 0$ and 
$z \frac{\partial^2 \gamma}{\partial H^2}|_{H=0} = -sgn(J)\beta^2 F$, 
where $F = 2^{-z+1}\sum_{i=1}^{z+1} Z_i ~^zC_{i-1}~ 
sech^2[\beta Z_i |J|]~tanh [ \beta Z_i |J|]$ with $Z_i = z - 2i + 2$. Here 
$sgn(J)$ is understood to be the sign of $J$, which is $+1$ for antiferromagnet
and $-1$ for ferromagnet. Now we write the Taylor series of $\tau_H^{-1}$ upto 
second order in $H$,
\begin{eqnarray}
\label{tau_H}
\tau_H^{-1} = \tau^{-1} -sgn(J)\alpha \beta^2 F ~H^2.
\end{eqnarray}
This relation shows that the inverse of relaxation time changes quadratically 
with (weak) magnetic field. For the ferromagnetic system, the relaxation time 
decreases with the strength of magnetic field while for the antiferromagnetic
system, it increases with the strength of magnetic field.

During second order phase transition, the relaxation time of a system is 
expected to diverge. Using this fact it is possible to estimate the critical 
temperature $T_C$ from Eq. (\ref{rlxt}). 
In the next section we study the criticality of the isotropic Ising 
systems in different dimensions.

\subsection{Study of criticality in different dimensions}
In this subsection, we study how one, two and three dimensional 
ferromagnetic isotropic Ising systems behave close to criticality. 
In particular, we calculate the critical 
temperatures ($T_C$) for the systems in different dimensions. 

For the isotropic case (when all coupling constant are same) all the 
parameters ($\gamma$'s) take the same value and is given by 
Eq. (\ref{istrgamma}) in 
the absence of magnetic field. Here the relaxation time of 
Eq. (\ref{rlxt}) will take the following form: 
$\tau = \frac{1}{\alpha}\left[1+z\gamma\right]^{-1}$. 
The critical temperature at which this 
relaxation time diverges can be found from the equation 
\begin{eqnarray}
\label{eqtc}
1+z\gamma = 0. 
\end{eqnarray}

For one dimensional isotropic system, $z = 2$ and 
$\gamma = -\frac 12 tanh~(2\beta |J|)$ (note for ferromagnetic systems, 
$J = - |J|$). In this case Eq. (\ref{eqtc}) takes the following form: 
$tanh~(2\beta |J|)=1$. This will be only satisfied when 
$\beta \rightarrow \infty$. Therefore in this case $T_C =0$, in accordance 
with the fact that the one dimensional Ising system behaves critically only 
near to absolute zero temperature.

For two dimensional system (square lattice), $z =4$ and 
$\gamma = - \frac 18 [ tanh~(4\beta |J|) + 2 tanh ~(2\beta |J|)]$. In this 
case Eq. (\ref{eqtc}) takes the following form: 
$tanh~(4\beta |J|) + 2 tanh ~(2\beta |J|) = 2$. Solution of this equation 
gives $T_C = 3.089|J|/k_B$, whereas its exact value is know to be 
$T_C = 2.269|J|/k_B$ \cite{kramers41,onsager44}. 

For three dimensional system (simple cubic lattice), $z = 6$ and 
$\gamma = - \frac{1}{32}[tanh~(6\beta|J|)+4 tanh~(4\beta |J|) 
+ 5 tanh ~(2\beta |J|)]$. In this 
case Eq. (\ref{eqtc}) takes the following form:
$[tanh~(6\beta|J|)+4 tanh~(4\beta |J|) + 5 tanh ~(2\beta |J|)]
=16/3$. Solution of this equation 
gives $T_C = 5.071|J|/k_B$, whereas its actual value is expected to be 
about $T_C = 4.511|J|/k_B$ \cite{salman98,livet91,talapov96}.

\begin{figure} []
\includegraphics[width=11.0cm]{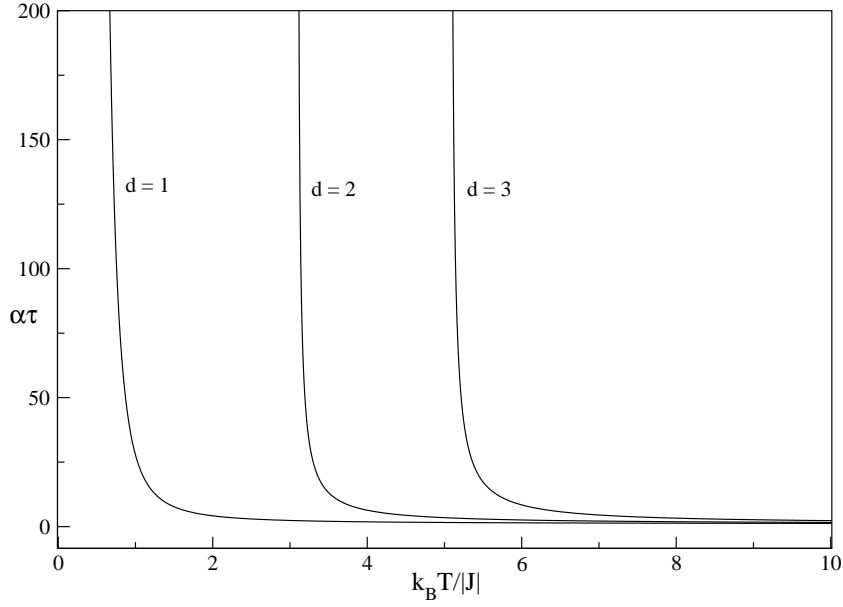}
\caption{Plots show how the relaxation time changes with 
temperature for three different dimensions.}
\label{tauplot}
\end{figure}
In Fig. \ref{tauplot}, one can see how the relaxation 
time $\tau$ changes with temperature; in particular, how it diverges near 
the criticality.

The values for the $T_C$ obtained for our optimal linear model are somewhat 
better in comparison with the mean field values (where $T_C = z|J|/k_B$ 
with $z$ = 2, 4 and 6 respectively for $d$ = 1, 2 and 3). 
Here it is encouraging to notice that our approach correctly captures the basic 
physics of the Ising model in different dimensions, viz., while the criticality 
exists only at absolute zero for one dimensional system, for two and three 
dimensional systems, the criticality exists at finite temperatures.

It may be worth mentioning here that, at criticality, $|\gamma| = \frac{1}{2d}$ 
(cf. Eq. (\ref{eqtc})), i.e., $\lambda = 1$ for the linear Glauber model 
(see Eq. (\ref{w_lgm})). This shows that our optimal linear model reduces to the 
voter model (without noise) at criticality ($T$ = $T_C$). This result gives 
a physical meaning to the fact that the linear Glauber model behaves critically 
when $\lambda = 1$.

\subsection{Static and dynamic scaling properties}
Analysis of our present optimal linear model essentially remains same as the 
linear Glauber model \cite{scheucher88,oliveira03,hase06}. We now discuss some 
of the important scaling properties of the linear model in the present context. 

To see how the static susceptibility ($\chi$) scales near critical temperature, 
we now find the expression for $\chi$. 
When $t \rightarrow \infty$, Eq. (\ref{magh}) gives, $\chi = \frac{\partial 
M(t \rightarrow \infty)}{\partial H}|_{H=0}  = 
- \alpha N \tau \frac{\partial \gamma_H}{\partial H}|_{H=0}$. Here $\tau$ is the 
relaxation time, which is given by Eq. (\ref{rlxt}). The parameter $\gamma_H$ 
is given by Eq. (\ref{istrgamma2_h}), from which we get 
$\frac{\partial \gamma_H}{\partial H}|_{H=0} = -\beta K$, where,
\begin{eqnarray}
\label{Kvalue}
K = 2^{-z} \sum_{i=1}^{z+1}~^zC_{i-1}~{\rm sech} ^2 [\beta Z_i J],
\end{eqnarray}
with $Z_i = z - 2i + 2$. At $T_C$, the term $K$ is finite, but $\tau$ 
diverges; this shows that the scaling behavior of $\chi$ will be same as 
$\tau$. To see how $\tau$ scales near criticality, we note that 
$\tau^{-1} = \alpha (1+z\gamma) = 0$ at criticality. Therefore, near criticality,  
$\tau^{-1} = \alpha |J|\beta_c R |\frac{T-T_C}{T_C}|$, with,
\begin{eqnarray}
\label{Rvalue}
R = 2^{-z} \sum_{i=1}^{z+1}Z_i^2 ~^zC_{i-1}~{\rm sech} ^2 [\beta Z_i J].
\end{eqnarray}  
To use in the expression of $\tau^{-1}$ near criticality, the term $R$ should 
be evaluated at $\beta = \beta_c$. It is now easy to see that, near criticality, 
$\chi \sim |\frac{T-T_C}{T_C}|^{-1}$ for all dimensions ($d \ge 2$). Therefore 
the critical exponent $\gamma$ = 1 (this symbol must not be confused with the 
optimization parameter $\gamma$).

To analyze the dynamic properties of the model, we now shift to the continuum 
limit. In this limit, the correlation function for two spins separated by 
the vector $\vec{r}$ satisfies the following {\it diffusion-decay} 
equation (this can be easily derived once we use the linear form of $w_k$'s 
in the equation of two-point correlation $r_{i,j}$; 
cf. Eq. (\ref{avr_t}) and Eq. (\ref{trate})):
\begin{eqnarray}
\label{dffeqn}
\frac{\partial G(\vec{r},t)}{\partial t} = D \nabla^2 G(\vec{r},t)
-\kappa G(\vec{r},t),
\end{eqnarray}
where $D = 2\alpha |\gamma|$ and $\kappa = 2\alpha (1-2d|\gamma|)$. 
We may note that, 
the solution of Eq. (\ref{dffeqn}) can be written in the following form,
\begin{eqnarray}
\label{crr_sln}
G(\vec{r},t) = P(\vec{r}) - Q(\vec{r},t)e^{-\kappa t},
\end{eqnarray}
where $P(\vec{r})$ and $Q(\vec{r},t)$ respectively satisfy the following 
equations: 
\begin{eqnarray}
\label{P_eq}
\nabla^2 P(\vec{r}) = \frac{\kappa}{D}  P(\vec{r}) ~~~{\rm and}\\
\label{Q_eq}
\frac{\partial Q(\vec{r},t)}{\partial t} = D \nabla^2 Q(\vec{r},t).
\end{eqnarray}

To solve  Eq. (\ref{dffeqn}), we must now set the initial condition(s) 
for $G(\vec{r},t)$ and discuss the asymptotic behavior of the function. 
We note that, $r_{i,i} = 1$ 
for all time $t$, also, $r_{i,j} = 0$ for $t = 0$ ($i \ne j$). In the 
continuum limit, it is convenient to set a lower cutoff $a > 0$, such 
that condition of self-correlation becomes, $G(|\vec{r}|=a,t) = 1$. 
Here the second condition becomes, $G(|\vec{r}|>a,t=0) = 0$. Physically, 
for large $|\vec{r}|$, the correlation between two spins will first 
increase and then saturate to its steady state value after a long time. 
The steady state value of the correlation function is given by 
$P(\vec{r})$ which is a solution of Eq. (\ref{P_eq}). It is here 
natural to set $P(a) = 1$. With the stated conditions on $G(\vec{r},t)$ 
and $P(\vec{r})$, we expect $Q(\vec{r},t)$  to behave in such a way that,
$Q(a,t) = 0$ and $Q(|\vec{r}|>a,0) = P(\vec{r})$ 
(cf. Eq. (\ref{crr_sln})). We now notice that, the solution for 
$Q(\vec{r},t)$ is the solution of a problem where we have an absorbing 
sphere of radius $a$ surrounded by moving particles. Initially, the 
concentration of the particles is high near the surface of the sphere and 
it decreases with the distance from the sphere according to the 
functional form of $P(\vec{r})$. Now following the same line of arguments 
as given in Ref.~\cite{krapivsky10}, it is possible to get the 
following asymptotic solution for $Q(r,t)$ (where $r = |\vec{r}|$; it 
should not be confused with the symbol for two-point correlation):
\begin{eqnarray}
\label{Q_sln}
Q(r,t) \simeq P(\sqrt{Dt}) \times \left\{ \begin{array}{l l l}
\frac{r}{\sqrt{Dt}},& d = 1 & \&~ 0< r \\
\frac{{\rm ln}(r/a)}{{\rm ln}(\sqrt{Dt}/a)}, & d = 2 & \&~ a< r \\
1-\frac{a}{r},&  d = 3 & \&~ a < r.
\end{array} \right.
\end{eqnarray}
It should be 
mentioned here that the functional form of $ P(\sqrt{Dt})$ will be 
different in different dimensions.
In Eq. (\ref{Q_sln}), we have assumed $r < \sqrt{Dt}$ for $d$ = 1 and 2. 
Accuracy of 
$Q(r,t)$ also depends on whether $\sqrt{Dt}$ is less than the correlation 
length ($\xi$). If $\sqrt{Dt} \gtrsim \xi $, then $Q(r,t)$ is expected 
to decay faster than what we get from Eq. (\ref{Q_sln}). This is 
because, the concentration of particles far away ($\gtrsim \xi$) from 
the absorbing sphere will be very low. In this situation, when 
$r \sim \xi \lesssim \sqrt{Dt}$, the particles at 
position $r$ will not only diffuse into the absorbing sphere, now they 
will also diffuse away towards the outer low concentration zone. 
For this reason, one expects Eq. (\ref{Q_sln}) to work 
good near criticality where correlation length is very large. 
We are here, though, not much interested in the case when the 
correlation length ($\xi$) is short; in such case, the correlation 
function $G(r,t)$ for large $r$ anyway always remains close to zero 
at all times. 

Now we turn our attention to find the steady state solution $P(r)$ from 
Eq. (\ref{P_eq}). A trial solution of the form $\frac{e^{-r/\xi}}{r^k}$ 
can be taken to find the desired solution for $P(r)$. Here 
$\xi = \sqrt{\frac{D}{\kappa}}$ is the correlation length and $k$ is a 
constant to be determined. We find that for $d$ = 1 and 3, $k$ = 0 and 1 
respectively. For $d = 2$, the above trial form does not yield any 
solution of Eq. (\ref{P_eq}). For this special case, we take the 
following trial form: $P(r) = S(r)e^{-r/\xi}$. If we put this in 
Eq. (\ref{P_eq}), we get the following equation for $S(r)$:
\begin{eqnarray}
\label{twosol}
\xi r \frac{d^2 S(r)}{dr^2} + (\xi -2r) \frac{dS(r)}{dr} - S(r) = 0.
\end{eqnarray}
The solution of this equation can be most easily found by a trial series 
of the form $S(r) = \sum_{n=0}^\infty a_n r^n$. After some calculations, 
we find that, $a_n = \frac{(2n-1)!!}{\xi^n(n!)^2}a_0$. This gives, 
$S(r) = a_0 [1+ \sum_{n=1}^\infty \frac{(2n-1)!!}{(n!)^2}(r/\xi)^n]$. By 
ratio test it can be verified that this series is convergent for any 
finite $r$. With this in hand, we now write solution for the $P(r)$:
\begin{eqnarray}
\label{P_sln}
P(r) = a_0e^{-r/\xi}\left\{ \begin{array}{l l}
1,& d = 1 \\
1+ \sum_{n=1}^\infty \frac{(2n-1)!!}{(n!)^2}(r/\xi)^n, & d = 2 \\
\frac{1}{r},&  d = 3.  
\end{array} \right.
\end{eqnarray}
The value of $a_0$ can be determined by the normalization condition 
$P(a) = 1$. For $d$ = 1 and 3, $a_0$ is respectively $e^{a/\xi}$ and 
$ae^{a/\xi}$. 
Near to the criticality ($\xi \rightarrow \infty$), 
it is easy to see that $P(r) \sim \mathcal{O}(1)$, $\mathcal{O}(1)$ and  
$1/r$ respectively for one, two and three dimension (we assume here 
$r \ll \xi$). This suggests that the critical exponent $\eta$, defined 
as $P(r) \sim r^{-(d-2+\eta)}{\rm e}^{-r/\xi}$, is 1, 
0 and 0 respectively for $d$ = 1, 2 and 3.

As we might expect, at criticality, the correlation function $G(r,t)$ 
reduces to the one for the voter model 
(cf. Ref.~\cite{krapivsky10}). This can be easily checked by 
taking $\xi \rightarrow \infty$ in Eqs. (\ref{Q_sln}) and 
(\ref{P_sln}). In this limiting case, the interface density $\rho$ (or 
domain wall density) scales with time as $t^{-1/2}$, $(ln~t)^{-1}$ and
$\mathcal{O}(1)$ respectively for $d$ = 1, 2 and 3. This temporal behavior 
is supposed to continue even when $T \gtrsim T_C$. This can be understood 
from the fact that $\rho$ is defined as $\frac 12 (1- G(r,t))$, where $r$ is 
the distance between nearest neighbors, i.e., technically, $r$ is here 
approximately equal to $a$ (but not exactly). If we now use Eqs. (\ref{Q_sln}) 
and (\ref{P_sln}) to calculate $\rho$, we will get back the same asymptotic 
temporal behavior as we have just mentioned. 
 
Before we calculate the dynamic exponent, let us first write 
the equation for the local magnetization $m(\vec{r},t)$ at the location 
$\vec{r}$ and time $t$
(this can be easily derived once we use the linear form of $w_k$'s
in the equation of $q_k$; cf. Eq. (\ref{avq_t}) and Eq. (\ref{trate})):
\begin{eqnarray}
\label{tdlg}
\frac{\partial m(\vec{r},t)}{\partial t} = \frac 12 D \nabla^2 m(\vec{r},t)
-\frac 12 \kappa m(\vec{r},t),
\end{eqnarray}
where again $D = 2\alpha |\gamma|$ and $\kappa = 2\alpha (1-2d|\gamma|)$. 
At criticality when $|\gamma| = \frac{1}{2d}$, we have $\kappa = 0$ and
$D = \alpha/d$; this reduces Eq. (\ref{tdlg}) to the diffusion
equation for the voter model \cite{krapivsky10}.
Eq. (\ref{tdlg}) is in the form of the well known time-dependent 
Ginzburg-Landau equation (linear version) for the non-conservative 
dynamics, which has been subject of active study for the last few 
decades \cite{puri09,krapivsky10,bray94}.
Advantage of our present work is that it gives us explicit temperature and 
exchange constant dependence of the parameters involved in the 
Ginzburg-Landau equation.
To gain some insight into Eq. (\ref{tdlg}), we will do a Fourier analysis of
the equation.
If we insert $m(\vec{r},t) = \sum_{\vec{k}} m(\vec{k},t) 
e^{i\vec{k}\cdot\vec{r}}$ in the equation, we get, 
\begin{eqnarray}
\label{franls}
\frac {\partial m(\vec{k},t) } {\partial t} = -\frac 12 (D k^2 +\kappa) 
m(\vec{k},t).
\end{eqnarray}
The solution of this equation gives,
\begin{eqnarray}
\label{mkt}
m(\vec{k},t) = m(\vec{k},0) e ^ {-t/\tau(k)},
\end{eqnarray}
where $\tau(k)$ is the relaxation time $(\frac{D}{2} k^2 +\frac{\kappa}{2})^{-1}$ 
for the $\vec{k}$ mode. This relaxation time can be rewritten as, 
\begin{eqnarray}
\label{rlxk}
\tau(k) = \frac{1}{\frac{D}{2}(k^2 +\xi^{-2})},
\end{eqnarray} 
where $\xi$ is, as mentioned earlier, the 
correlation length:  $\xi = \sqrt{D/\kappa} = 
\sqrt{\frac{|\gamma|}{1-2d|\gamma|}}$. Here it may be briefly mentioned that, 
near criticality the correlation length diverges as,  
$\xi \sim |\frac{T-T_C}{T_C}|^{-1/2}$. 
This shows that the critical exponent $\nu = 1/2$ (for $d$ = 2 and 3).

The dynamic exponent (denoted by $z$; not to be confused 
with coordination number) is defined by how the maximum possible value of 
the relaxation time ($\tau_{\rm max}$) scales with the system's relevant length 
scale. For a thermodynamic system (size $L \rightarrow \infty$) where lowest 
possible value of $k$ is zero, we get from Eq. (\ref{rlxk}), 
$\tau_{\rm max} \sim \xi^2$, i.e., the exponent $z = 2$. On the other hand, 
for a finite system of size $L$, near criticality, 
$\tau_{\rm max} \sim k_{\rm min}^2$, i.e., $\tau_{\rm max} \sim L^2$. Here 
again the exponent $z = 2$.

\subsection{Fluctuation-dissipation theorem}
In this subsection we will discuss the fluctuation-dissipation theorem (FDT) for 
our optimal linear model. First we will calculate the dynamical susceptibility, 
$\chi(\omega)$, then we will establish its relation to the autocorrelation of 
the total stochastic magnetization function. 

Before we calculate $\chi(\omega)$, we first assume that the magnetic field 
dependent transition rate $w'_j$ and Eq. (\ref{magh_eq}) are valid even when 
the magnetic field ($H$) is time-dependent. We also assume that the field is 
weak, i.e., $|H| \ll k_BT$ and the system is above critical temperature 
($T > T_C$). 

Noting the fact that $\tau_H \simeq \tau$ in the first order approximation 
(since, $|\tau_H - \tau|\sim |H|^2$; cf. Eq. (\ref{tau_H})), 
we recast Eq. (\ref{magh_eq}), in the following way,
\begin{eqnarray}
\label{mtime_eq}
\frac{d}{dt}M(t) =-\frac{1}{\tau}M(t) -\alpha N \gamma_H. 
\end{eqnarray}
Assuming the system was in the steady state 
before the weak time-dependent magnetic field $H(t)$ was applied, the complementary
solution of the nonhomogeneous first order linear differential Eq. (\ref{mtime_eq}) 
will be zero while the particular solution of the equation will give us the solution 
for $M(t)$. Assuming that $H(t)$ was applied in the distant past 
($t\rightarrow -\infty$), we have the following particular solution, 
\begin{eqnarray}
\label{mtime}
M(t) = - \alpha N \int_{-\infty}^t \gamma_H(t')e^{-\frac{(t-t')}{\tau}}dt'.
\end{eqnarray}
Now we recognize that, $\gamma_H(t)= - \beta K H(t)$, where $K$ is given by 
Eq. (\ref{Kvalue}). Assuming $H(t) = H_0 e^{-i\omega t}$, we get from 
Eq. (\ref{mtime}), $M(t) = \chi(\omega) H_0 e^{-i\omega t}$, where the dynamical 
susceptibility is given by,
\begin{eqnarray}
\label{dynchi}
\chi(\omega) = \frac{\alpha N K \beta}{1/\tau - i\omega}.
\end{eqnarray}
In the low frequency limit, $\omega \rightarrow 0$, we get back the static 
susceptibility, $\chi = \alpha N K \beta \tau$ (see preceding subsection). 

To verify the FDT for our model, we now calculate the autocorrelation of the 
total stochastic magnetization function, 
\begin{eqnarray}
\label{autocrr}
<\Sigma(0)\Sigma(t)> = \sum_{\{\sigma\},\{\sigma'\}} 
p(\{\sigma\};0)\Sigma(0)p(\{\sigma'\};t|\{\sigma\};0)\Sigma(t),
\end{eqnarray}
where $\Sigma(0) = \sum_{i=1}^N \sigma_i(0)$ and $\Sigma(t) = 
\sum_{i=1}^N \sigma'_i(t)$.
Here $p(\{\sigma'\};t|\{\sigma\};0)$ is the conditional probability that the 
total stochastic magnetization function will assume the value $\Sigma(t)$ at 
time $t$ if it initially assumes the value $\Sigma(0)$ with the probability 
$p(\{\sigma\};0)$. We now note that, 
$ \sum_{\{\sigma'\}} p(\{\sigma'\};t|\{\sigma\};0)\Sigma(t) 
= M(0) e^{-t/\tau}$ 
(cf. Eq. (\ref{mag})). Here $M(0)$ is understood to be the initial value of 
the stochastic magnetization function, i.e., $M(0)=\Sigma(0)$. This allows us 
to rewrite Eq. (\ref{autocrr}) as, 
$<\Sigma(0)\Sigma(t)> = \sum_{\{\sigma\}} 
P(\{\sigma\};0)\Sigma^2(0)e^{-|t|/\tau}$. We now see that, $\sum_{\{\sigma\}} 
P(\{\sigma\};0)\Sigma^2(0) = \sum_{i,j=1}^N r_{i,j}$, which is $\chi/\beta$ 
in the steady state (this is because the total magnetization is zero in the 
steady state). In the next step we do a Fourier 
transform of the autocorrelation function,
\begin{eqnarray}
\label{atcrrFT}
\int_{-\infty}^{\infty} <\Sigma(0)\Sigma(t)> e^{i\omega t} dt &=& 
\frac{2\alpha N K}{1/\tau^2+\omega^2} \\ \nonumber
&=& \frac{2 k_B T}{\omega} ~{\rm Im}~ \chi(\omega).
\end{eqnarray}
This shows that the FDT is valid for our optimal linear model, i.e., 
the Fourier transform of the autocorrelation function is proportional to the 
dissipative part (or imaginary part) of the the dynamical susceptibility. 

\section{Conclusion}
In this paper we propose a new analytical method to study the Glauber dynamics 
in an arbitrary Ising system (in any dimension). 
It is know that, unlike its nonlinear version, the linear Glauber model (LGM) 
is exactly solvable even though the detailed balance condition is not generally 
satisfied. Motivated by the fact, 
we have here addressed the issue of writing the transition rate ($w_j$) in
a best possible linear form such that the mean squared error in satisfying the
detailed balance condition is least. 
This serves the following purpose:
by studying the LGM analytically, we will be able to anticipate how the
kinetic properties of an arbitrary Ising system
depend on the temperature and the coupling constants.
For a generic system, we have shown how this optimization can be done 
using a simple Moore-Penrose pseudoinverse matrix. 
This approach is quite general, applicable to arbitrary system and can
reproduce the exact results for one dimensional Ising system. From this 
perspective, our work can be viewed as the generalization of Glauber's 
work for one dimensional Ising system.
In the continuum limit, our approach leads to a linear time-dependent 
Ginzburg-Landau (TDGL) equation of non-conservative dynamics. This
establishes a connection between the phenomenological TDGL theory and the
Glauber's microscopic model for non-conservative dynamics. 
Both the static (steady state) and dynamic properties of the Ising systems (in 
different dimensions) are analyzed using our optimal linearization approach.  
We saw that most of the important results  
obtained in different studies can be reproduced by our new mathematical approach.

We also demonstrated in our paper that the effect of the magnetic field can be 
easily treated within our approach; our transition rate in the presence of 
magnetic field works more efficiently than the commonly used one. In particular, 
we showed that the fluctuation-dissipation theorem is valid for our optimal 
linear model and that the inverse of relaxation time changes quadratically with 
the applied (weak) magnetic field. 

We hope that our present mathematical approach can also be extended to study the 
microscopic dynamics in other systems like the Potts model, Heisenberg model, etc. 
It should 
also be useful to study other kinetic models. It may be mentioned here that this 
approach has already been used to study the Kawasaki model for conservative 
dynamics \cite{sahoo14}.

\begin{acknowledgements}
SS thanks Prof. S. Ramasesha for his financial support through his various 
projects from IFCPAR and DST, India, and SKG thanks CSIR, India for financial 
support.
\end{acknowledgements}

\end{document}